\documentclass[12pt]{article}
\usepackage{amsthm,amsfonts,amsmath, amscd, bbold}
\usepackage{mathrsfs}
\usepackage{authblk}

                                \usepackage{verbatim}
\swapnumbers
                              
\theoremstyle{plain}

\usepackage{enumitem}
\numberwithin{equation}{section}
\newtheorem{theorem}{Theorem}[section]

\newtheorem{assumption}[theorem]{Assumption}

\theoremstyle{definition}

\theoremstyle{remark}

\numberwithin{equation}{section}

\newcommand{\cE}{{\mathcal E}}
\newcommand{\cH}{{\mathcal H}}

\newcommand{\ket}[1]{\left\vert #1\right\rangle}
\newcommand{\bra}[1]{\left\langle #1\right\vert}

\newcommand{\Tr}{\mathrm{Tr}}

\newcommand{\be}{\begin{equation}}
\newcommand{\ee}{\end{equation}}
\newcommand{\bea}{\begin{eqnarray}}
\newcommand{\eea}{\end{eqnarray}}
\newcommand{\beann}{\begin{eqnarray*}}
\newcommand{\eeann}{\end{eqnarray*}}

\usepackage{color}

\begin{document}

\title{Entanglement rates for Renyi, Tsallis and other entropies}
\author{Anna Vershynina\\
\small{\it Department of Mathematics, Philip Guthrie Hoffman Hall, University of Houston,}\\ 
\small{\it 3551 Cullen Blvd., Houston, TX 77204-3008, USA}\\
\small{annavershynina@gmail.com}}

\date{ }

\maketitle

\begin{abstract}
We provide an upper bound on the maximal entropy rate at which the entropy of the expected density operator of a given ensemble of two states changes under nonlocal unitary evolution. A large class of entropy measures in considered, which includes R\'enyi and Tsallis entropies. The result is derived from a general bound on the trace-norm of a commutator, which can be expected to find other implementations. We apply this result to bound the maximal rate at which quantum dynamics can generate entanglement in a bipartite closed system with R\'enyi and Tsallis entanglement entropy taken as measures of entanglement in the system. \end{abstract}

\section{Introduction}

The problem addressed in this paper is, given a bipartite system between two parties, Alice and Bob, to find an upper bound on the rate at which entanglement can be generated in time. Different entanglement measures can be considered, for example R\'enyi and Tsallis entanglement entropies are obtained as a particular case of a more general result.

Let us say that Alice and Bob have access to systems $A$ and $B$ respectively together with local ancilla systems $a$ and $b$ respectively. The system starts in a pure state $\ket{\Psi}_{aABb}$ and evolved according to a Hamiltonian $H_{AB}$ that acts only on systems $A$ and $B$. Since the state of the system $\ket{\Psi(t)}_{aABa}=\exp\{it H_{AB}\}\ket{\Psi}_{aABa}$ stays always pure, one may calculate the entanglement in the system $E(t)$ using various entanglement measures, e.g. von Neumann entanglement entropy, R\'enyi entanglement entropy or Tsallis entanglement entropy. The {\it entanglement rate} is a time derivative of the entanglement measure at time $t=0$, $\Gamma(\Phi, H)=dE(t)/dt|_{t=0}$. The small incremental entangling problem aims at finding an upper bound on the maximal entanglement rate $\Gamma(H)=\sup_{\ket{\Psi}_{aABb}}\Gamma(\Psi, H)$ that is independent of dimensions of ancillas and the initial state $\ket{\Psi}_{aABb}$. The problem of maximizing the entangling rate of a bipartite system in the presence of ancillas evolving under unitary dynamics was studied by many authors, but only for the case of a von Neumann entanglement entropy.

In a case when $A$ and $B$ are qubits, Childs et al \cite{CLVV03} gave upper bounds for an entangling rate and showed that they are independent of ancillas $a$ and $b$. Wang and Sanders \cite{WS03} considered systems $A$, $B$ and ancillas of arbitrary dimensions and proved that the entangling rate is upper bounded by $\Gamma(H)\leq\beta\approx1.9123$, for a self-inverse product Hamiltonian. Bravyi \cite{B07} proved that in a general case with no ancillas the entangling rate is bounded by $\Gamma(H)\leq c(d)\|H\|\ln d$, where $d=\min(d_A, d_B)$ is the dimension of the interacting subsystems and $c$ is a constant close to 1. For an arbitrary bipartite Hamiltonian Bennett et al \cite{BHLS03} proved that the upper bound on the ancilla-assisted entanglement is independent of the ancilla dimensions $\Gamma(H)\leq c d^4\|H\|$, where $c$ does not depend on $a$ or $b$. The bound was improved by Lieb and Vershynina \cite{LV13} providing an upper bound $\Gamma(H)\leq 4\|H\|\ln d$ for an arbitrary Hamiltonian in ancilla-assisted system. Finally the question was answered by Van Acoleyen et al. \cite{AMV13} arriving at $\Gamma(H)\leq 18\|H\|\ln d$. Few months later an independent proof was presented by Audenaert \cite{A13} that gives an upper bound $\Gamma(H)\leq 8\|H\|\ln d$. In \cite{N16} Ning et. al. improved the constant to the one believed to be the most optimal for a specific case of states $\Gamma(H)\leq 2\|H\|\ln d$. The bound with logarithmic dependence is optimal, since one can find a particular Hamiltonian $H_{AB}$ for which they there is an equality. See Marien et al. \cite{MAVV13} for a detailed review on the entangling rates for bipartite closed systems, which we will also quickly review in the next section. 

The same question was also posed by Vershynina \cite{V15} for an open system, when the generator of the dynamics consists of both Hamiltonian and dissipative part. In that paper the relative entropy of entanglement was chosen as a measure of entanglement in an ancilla-free system, and the author provided an upper bound on the entangling rate, which has a logarithmic dependence on a dimension of a smaller system in a bipartite
cut. The rate of change of quantum mutual information in an ancilla-assisted system was also investigated, and it was shown that an upper bound is independent of dimension of ancillas.

Here we generalize previous results in the closed systems to include not only von Neumann entanglement entropy, but R\'enyi, Tsallis, and a wide class of entanglement measures, see (\ref{eq:ent-measure2}).

In most of the works mentioned above the bound on the entanglement rate is obtained as a consequence of a more general problem: small incremental mixing. The problem is, given a probabilistic ensemble of states $\cE=\{(p,\rho_1), (1-p, \rho_2)\}$, to find an upper bound on the rate, $\Lambda(\cE)$, at which the von Neumann entropy of the expected  density operator of this ensemble changes under a non-local evolution. We review this problem and its relation to the entanglement rate problem in detail in the next section. To bound $\Lambda(\cE)$ one  needs to find an upper bound on the following quantity 
\begin{equation}\label{eq:intro}
\|[X, f(Y)]\|_1,
\end{equation}
where $f(t)=\ln(t)$, $0\leq X\leq Y\leq I$ with $\Tr X=p$ and $\Tr Y=1$. Obtaining the upper bound in the form $-p\ln p$ would result in the logarithmic upper bound in the entanglement rate problem. It's worth mentioning that for this problem a more optimal bound for $1/100<p<1-1/100$ was obtained by Lieb and Vershynina \cite{LV13}, providing $4\sqrt{p(1-p)}$ as an upper bound. Here $p$ can be viewed as the inverse of the dimension $d$, so for the entanglement rate problem we are interested in the case when $d$ is big, i.e. $p$ is small.

Audenaert and Kittaneh \cite{AK12} posed a question of bounding expression (\ref{eq:intro}) for a general function $f$. Their conjecture was that for some class of functions $f$ the following holds
\begin{equation}\label{eq:intro2}
\|[X, f(Y)])\|_1\leq c(F(1)-F(p)-F(1-p)),
\end{equation}
where $F(q)=\int_0^q f(t)dt$, and operators $X$ and $Y$ satisfy conditions above.

We provide a different upper bound on the expression (\ref{eq:intro}) for a large class of functions, arriving at
\begin{equation}\label{eq:intro3}
\|[X, f(Y)])\|_1\leq 9 \min\{ p (f(1)-f(p)), (1-p)(f(1)-f(1-p))\}.
\end{equation}
In section 3 we compare the conjecture by Audenaert and Kittaneh with the bound derived, and show that for the majority of cases the bound we provide gives better scaling for small $p$.

The paper is organized as follows. In Section 2 we discuss Small Incremental Entangling  (SIE) problem, Small Incremental Mixing (SIM) problem,  and the relation between the two. In Section 3 we consider a Generalized SIM problem, and provide a main result Theorem \ref{thm:main} on its upper bound. At the end of the section we compare our result with the conjecture in \cite{AK12}. In Section 4 we apply the main result to obtain the bound on the R\'enyi entanglement entropy. In Section 5 we apply the main result for the entropy rate for Tsallis entropy. In Section 6 we prove the main result Theorem \ref{thm:main}. In Section 7 we prove that the power function is included in the class of functions for which Theorem \ref{thm:main} holds. Section 8 is reserved for the conclusion of the paper.

\section{Preliminaries}\label{sec:prelim}

Our main result in Theorem \ref{thm:main} is motivated by the question of bounding the rate at which entanglement changes in quantum system under unitary evolution. In Section \ref{sec:SIE} we discuss this problem, known as Small Incremental Entangling (SIE). A more general problem, known as Small Incremental Mixing (SIM) can be used to prove SIE and is discussed in Section \ref{sec:SIM}. The relation between two problems is shown in Section \ref{sec:SIE-SIM}.

\subsection{Small Incremental Entangling}\label{sec:SIE}

Suppose that two parties, say Alice and Bob, have control over systems $A$ and $B$ and ancillary systems $a$ and $b$, respectively. The systems $A$ and $B$ evolve according to a non-local Hamiltonian $H_{AB}$. 

 Alice and Bob start with a pure state $\rho(0)=\ket{\Psi}\bra{\Psi}$ on the system $aABb$.
A time dependent joint state of Alice and Bob is a pure state
$$\rho(t)=U^*(t)\ket{\Psi}\bra{\Psi}U(t),$$
 where $U(t)=I_a\otimes e^{iH_{AB}t}\otimes I_b$
is a unitary transformation. 

One of the ways to describe the entanglement between Alice and Bob is to calculate the entanglement entropy
$$E(\rho(t)):=S(\rho_{aA}(t))=-\Tr\rho_{aA}(t)\ln \rho_{aA}(t),$$
where $\rho_{aA}(t)=\Tr_{Bb}\rho(t)$ is a state that Alice has after time $t$ and $S(\rho)$ is a von Neumann entropy. Since the joint state is pure, the entanglement entropy
 can also be calculated from the Bob's state $E(\rho(t))=S(\rho_{Bb}(t))$.

In 2003 Bennet et. al. \cite{BHLS03} proved that the total change of entanglement has the following upper bound, see \cite{MAVV13} for a simpler and more intuitive proof.

\textbf{Small Total Entangling.}\textit{ The total change of the entanglement $E(\rho(t))$ is at most $2\ln d$, where $d=\min\{\dim(A), \dim(B)\}$.} 

A question of bounding the infinitesimal change of the entanglement is formulated using the entangling rate. The \textit{{entangling rate}} is defined as a derivate of the entanglement entropy
$$ \Gamma(\Psi, H)=\frac{dE(\rho(t)}{dt}\bigg|_{t=0}.$$
After calculating the derivative, the entangling rate can be expressed as
\begin{align}
 \Gamma(\Psi, H)&=-i\Tr\Bigl(H_{AB}[\rho_{aAB}, \ln(\rho_{aA})\otimes I_B]\Bigr)\nonumber\\
&=-i\Tr\Bigl(H_{AB}[\rho_{aAB}, \ln(\rho_{aA}\otimes \frac{I_B}{\dim(B)})]\Bigr).\label{Ent_rate}
\end{align}

The following statement was conjectured by Bravyj \cite{B07} with constant $2$, proved by Van Acoleyen et. al. \cite{AMV13}, and implicitly improved by Audenaert \cite{A13} providing constant {$4$}. For a specific class of states Ning et. al. \cite{N16} improved the constant to $2$.

\begin{theorem}\label{thm:SIE}(\cite{AMV13}) \textbf{Small Incremental Entangling.}\\
 For all dimensions of ancillas $a$, $b$ and for all initial states $\ket{\Psi}$, the 
entanglement rate is bounded by
$$\Gamma(\Psi, H)\leq 18 \|H\|\ln d, $$
{where $d=\min\{\dim(A), \dim(B)\}.$}
\end{theorem}

The Small Incremental Entangling (SIE) problem is closely related to a more general Small Incremental Mixing (SIM) problem. The proof of SIE presented by Van Acoleyen et. al. in \cite{AMV13} relies on the proof of SIM. In the next section we discuss SIM problem and in Section \ref{sec:SIE-SIM} we show how to prove SIE having proved SIM.

\subsection{Small Incremental Mixing}\label{sec:SIM}

Let $\cH$ be a $D$-dimensional Hilbert space. The Hilbert space could be infinite dimensional, see \cite{MAVV13}, but in this paper will be assumed to be finite dimensional. Our result is expected to be carried out in the infinite-dimensional set up as well by following the reasoning in \cite{MAVV13}.

 Let $\cE_2=\{(p,\rho_1), (1-p, \rho_2)\}$ be a probabilistic ensemble of two states acting
on $\cH$. 
The expected density operator of this ensemble is a convex combination $\rho=p\rho_1+(1-p)\rho_2$. For any Hamiltonian $H$ (self-adjoint
operator on $\cH$) define
a time dependent state $$\rho(t)=p\rho_1+(1-p)e^{-iHt}\rho_2e^{iHt}.$$

The von Neumann entropy of this state is $$S(\rho(t))=-\Tr\Bigl(\rho(t)\ln\rho(t)\Bigr).$$

From the basic properties of von Neumann entropy the following statement holds, see \cite{LV13} for a detailed proof.

\textbf{Small Total Mixing.}\textit{ For any fixed ensemble $\cE_2$, the entropy of a state $\rho(t)$ at any time $t$ satisfies
$$ \overline{S}(\cE_2)\leq S(\rho(t))\leq \overline{S}(\cE_2)+S(p),$$
where $\overline{S}(\cE_2)=pS(\rho_1)+(1-p)S(\rho_2)$ is the average entropy of the ensemble and 
$S(p)=-p\ln p-(1-p)\ln(1-p)$ is a binary entropy.}

The  analogue of the small total mixing for infinitely small times is formulated in terms of 
a mixing rate.
A \textit{{mixing rate}} $\Lambda$ is defined as
\begin{equation}\label{eq:def-mixing}
\Lambda=|\Lambda(H)|, \text { where } \Lambda(H)=\frac{dS(\rho(t))}{dt}\bigg|_{t=0}.
\end{equation}
Calculating the derivative of the entropy we find that
\begin{align}\label{L_bin}
 \Lambda(H)&=-ip\Tr([\rho_1,\ln\rho]H)\\
&=i(1-p)\Tr([\rho_2,\ln\rho]H)\nonumber.
\end{align}
Denoting $X:=p\rho_1$ and $Y:=\rho$ we obtain
\begin{equation}
\Lambda=|\Tr(H [X, \ln (Y)])|,
\end{equation}
here $0<X\leq Y\leq I$ with $\Tr(X)=p$, $\Tr(Y)=1$.

Using that a trace-norm can be written as $\|Z\|_1=\Tr|Z|=\max_{\|H\|=1}(HZ)$ we have
\begin{align}
 \Lambda&\leq \Tr|[X,\ln(Y)]|\\
&=\|[X,\ln(Y)]\|_1,
\end{align}
here the maximum is achieved for $H=1-2R$, with $R$ being a projector on the negative eigenspace of $i[X,\ln(Y)]$.

The following statement was conjectured by Bravyi \cite{B07} with constant $1$, proved by Van Acoleyen et. al. \cite{AMV13}, improved by Audenaert \cite{A13} with constant $2$, and proved by Ning et al. \cite{N16} with constant $1$ for a specific class of states.

\begin{theorem}\label{thm:SIM}\textbf{Small Incremental Mixing}.\\
{For any ensemble $\cE_2=\{(p,\rho_1), (1-p, \rho_2)\}$, the maximum mixing rate is bounded above by a binary entropy}
\begin{align}
 \Lambda=\|[X,\ln(Y)]\|_1&\leq 9 S(p)=9(-p\ln p-(1-p)\ln(1-p)).\nonumber
\end{align}
\end{theorem}

In fact, Van Acoleyen et. al. \cite{AMV13} proved the following stronger statement.
For  $0<p\leq 1/2$, the maximum mixing rate is bounded above by
\begin{align}
\|[X,\ln(Y)]\|_1&\leq -9 p\ln p.\label{eq:SIM2}
\end{align}

For $1/100<p<1-1/100$ an even stronger bound on the mixing rate was proved by Lieb and Vershynina \cite{LV13}.  
$$  \|[X,\ln(Y)]\|_1\leq 4\sqrt{p(1-p)}.$$

\subsection{SIE follows from SIM}\label{sec:SIE-SIM}

Bravyi \cite{B07} proved that Small Incremental Mixing, Theorem \ref{thm:SIM}, with a constant $c$ in front of the Shannon entropy implies Small Incremental 
Entangling, Theorem \ref{thm:SIE}, with a constant $2c$.

In Lemma 1 \cite{B07} Bravyi proved that there exists a state $\mu_{aAB}$, such that a state appearing  in (\ref{Ent_rate}) can be written as 
\begin{equation}
\rho_{aA}\otimes\frac{I_B}{B}=\Bigl(1-\dim(B)^{-2}\Bigr)\mu_{aAB}+\dim(B)^{-2}\rho_{aAB}.
\end{equation}
This state is also an expected density operator of the following ensemble of states $$\cE_2=\{((1-p), \mu_{aAB}), (p, \rho_{aAB})\}$$ with $p=\dim(B)^{-2}$. Here without loss of generality it was assumed that $\dim(B)\leq \dim(A)$.

Applying the expression for the mixing rate (\ref{L_bin}) to this ensemble, we have that $$\Lambda( H)=p\Gamma(\Psi, H)=\dim(B)^{-2}\Gamma(\Psi, H),$$ which shows that Small
Incremental Mixing implies Small Incremental Entangling with the increase of the constant.

\section{Generalized SIM}\label{sec:Gen}

In the previous sections we considered von Neumann entropy as the entropy measure. But one may consider other entropy measures and ask the same question of bounding the mixing rate for such an entropy measure. And even more generally, instead of a logarithmic function in the commutator appearing in Theorem \ref{thm:SIM} one may consider any function $f$. In other words, we investigate  the expression
\begin{equation} \label{eq:expression}
\|[X, f(Y)])\|_1,
\end{equation}
for $0\leq X\leq Y\leq I$ with $\Tr X=p$, $\Tr Y=1$.

In 2012 Audenaert  and Kittaneh \cite{AK12} were the first to consider this expression, and for some class of functions $f$ they conjecture that
\begin{equation}\label{eq:Aud-conj}
\|[X, f(Y)])\|_1\leq c(F(1)-F(p)-F(1-p)),
\end{equation}
where $F(q)=\int_0^q f(t)dt$, and operators $X$ and $Y$ satisfy conditions above.

Here we provide a different upper bound on (\ref{eq:expression}) for a class of function satisfying the following condition on their growth.

\begin{assumption}\label{Assum:func}
For $0<p<1$ and a function $f:[0,1]\rightarrow \mathbb{R}$ assume that
\begin{enumerate}
\item $f$ is monotonically increasing;
\item for $0<y<x\leq 1$,
\begin{align}
\text{if } f(x)-f(y)&>f(1)-f(p),\label{eq:ass1}\\
\text{then } x^{-1/2}y^{1/2}(f(x)-f(y)) &\leq p^{1/2}(f(1)-f(p)).\label{eq:ass2}
\end{align}
\end{enumerate}
\end{assumption}

In the Section \ref{sec:Func} we show that function $f(t)=t^{\beta}$, $\beta>0$ satisfies this Assumption for small enough $p$, and function $f(t)=-t^{\beta}$, $\beta<0$ satisfies this Assumption for all $p$. The logarithmic function $f(t)=\ln(t)$ satisfies this assumption which led to the first proof of the small incremental problem in \cite{AMV13}. Here we show that a similar proof holds for a power function, which appears in the Renyi entropy, and all functions satisfying  Assumption \ref{Assum:func}.

Clearly the function can be taken to be monotonically decreasing in the Assumption, then the function $-f$ that satisfies  Assumption \ref{Assum:func} will be taken for the generalized small incremental problem (\ref{eq:expression}) without changing it.

Note that if the function $f$ has an inverse function $\mathring{f}$, the statement (\ref{eq:ass2}) is a weaker version of a monotonicity statement for a function $g$ specified below. Since $f$ is monotonically increasing, from (\ref{eq:ass1}) we get
$$x>\mathring{f}(f(y)+f(1)-f(p)).$$
If  $\mathring{f}(f(y)+f(1)-f(p))>0$, then, since $x\leq 1$, $f(x)\leq f(1)$, and the left-hand side of (\ref{eq:ass2}) is not greater than the following function \begin{equation}\label{eq:monotone-function}
g(y):=\left\{ \mathring{f}(f(y)+f(1)-f(p))\right\}^{-1/2} y^{1/2}(f(1)-f(y)).
\end{equation}
The right-hand side of (\ref{eq:ass2}) is equal to $g(p)$. Note that from (\ref{eq:ass1}) we have that $0<y<p$, because otherwise we would have $p\leq y<x\leq 1$, which would lead to the failure of  (\ref{eq:ass1}). Therefore, if the function $g$ is proven to be monotonically increasing, the statement (\ref{eq:ass2}) follows. 

The main result of this paper is the following upper bound on (\ref{eq:expression}) for functions satisfying Assumption \ref{Assum:func}.

\begin{theorem}\label{thm:main}
 Let function $f$ satisfy Assumption \ref{Assum:func} for $0<p\leq 1$. Then the trace-norm of the commutator is bounded above by 
\begin{equation}\label{eq:thm-main}
\|[X, f(Y)])\|_1\leq 9 \min\{ p (f(1)-f(p)), (1-p)(f(1)-f(1-p))\},
\end{equation}
for $0\leq X\leq Y\leq I$, $\Tr X=p$, $\Tr Y=1$.
\end{theorem}
The proof of this theorem is given in Section \ref{sec:Proof}.

Note that the right-hand side of  (\ref{eq:thm-main}) is symmetric in $p$ and $1-p$ since $X$ can be replaced with $Y-X$ in the trace-norm. 

This theorem can be formulated in the infinite-dimensional space, with the proof following \cite{MAVV13} leading to the increase of the constant. Since the essence of both proofs is the similar, we will restrict ourselves to a finite-dimensional setup. 

Let us compare the upper bound (\ref{eq:thm-main}) (without constant $9$) with bound (\ref{eq:Aud-conj}), conjectured by Audenaert and Kittaneh \cite{AK12}, for several functions:

\begin{itemize}
\item $f(t)=\ln(t)$: the upper bound (\ref{eq:thm-main}) is clearly more tight than (\ref{eq:Aud-conj}). The bound (\ref{eq:thm-main}) is exactly the same as was proved for SIM (\ref{eq:SIM2}), and the bound (\ref{eq:Aud-conj}) is the same as originally conjectured Theorem \ref{thm:SIM};
\item $f(t)=t^\beta$, $\beta\geq 1$ and {small enough $p$}: the conjectured bound (\ref{eq:Aud-conj}) gives a better scaling than (\ref{eq:thm-main}). But note that derivatives of the right-hand sides of (\ref{eq:thm-main}) and  (\ref{eq:Aud-conj}) at $p=0$ are the same and are equal to one.
\item $f(t)=t^\beta$, $0<\beta\leq 1$ and {small enough $p$}:  bound (\ref{eq:thm-main}) gives a better scaling than  the conjectured bound (\ref{eq:Aud-conj}). The derivative of the right-hand side of (\ref{eq:thm-main}) at $p=0$ is $\beta$, while the derivative of the right-hand side of (\ref{eq:Aud-conj}) at $p=0$ is one, which is not smaller than $\beta$.
\end{itemize}

For the above function $f(t)=t^\beta$ and appropriate $p\leq 1/2$, Theorem \ref{thm:main} takes the following explicit form:
\begin{enumerate}[label={P.\arabic*.},ref=P.\arabic*]
\item\label{power-thm1} $1\leq\beta$: we have $\|[X, Y^\beta]\|_1\leq 9 p(1-p^\beta)$,
\item\label{power-thm2}   $0\leq\beta<1$: we have $\|[X, Y^\beta]\|_1\leq 9 (1-p)(1-(1-p)^\beta)$.
\end{enumerate}

The expression $\|[X, f(Y)])\|_1$ can be viewed as a mixing rate for any entropy of the form
\begin{equation}\label{eq:ent-measure}
S(\rho)=k_1\Tr(F(\rho))+k_2;
\end{equation}
with function $F(x)$ such that $F'(x)=f(x)+const$ and $k_1,k_2\in\mathbb{R}$. The mixing rate (\ref{eq:def-mixing}) for this entropy is then given by
\begin{equation}\label{eq:ent-rate}
\Lambda(H)=k_1\Tr (H[X, f(Y)]),
\end{equation}
with $X$, $Y$ and $H$ discussed in Section \ref{sec:SIM}.
The von Neumann entropy $S(\rho)=-\Tr(\rho\log \rho)$,
discussed in the previous section, is an example of such an entropy. A Tsallis-$q$ entropy \cite{AD75, D70} is another example of such a measure. It is defined as
\begin{equation}\label{eq:Tsallis}
S_q(\rho)=\frac{1}{q-1}(1-\Tr(\rho^q)),
\end{equation}
for $q>0$, $q\neq 1$. Quantum Tsallis-$q$ entropy is a generalization of von Neumann entropy with respect to the parameter $q$, i.e. Tsallis-$q$ entropy converges to von Neumann entropy when $q$ tends to $1$,
\begin{equation}
\lim_{q\rightarrow 1}S_q(\rho)=S(\rho).
\end{equation}
The entanglement rate problem for Tsallis entropy is discussed in the Section \ref{sec:Tsallis}.

Moreover, for any entropy  of the form
\begin{equation}\label{eq:ent-measure2}
S(\rho)=k_1\log\Tr(F(\rho))+k_2,
\end{equation}
with function $F(x)$ such that $F'(x)=f(x)+const$, and $k_1,k_2\in\mathbb{R}$ the entropy rate is given by
\begin{equation}\label{eq:ent-rate2}
\Lambda(H)=\frac{k_1}{\Tr(F(Y))}\Tr (H[X, f(Y)]).
\end{equation}
Provided that $\Tr(F(Y))$ can be properly bounded one can bound the entropy rate using (\ref{eq:thm-main}).
The Renyi entropy 
$$S_\alpha(\rho)=\frac{1}{1-\alpha}\log \Tr (\rho^\alpha), \qquad \text{with }\alpha\geq 0,$$
is an example of such an entanglement measure. Note that Renyi entropy is another generalization of von Neumann entropy, i.e. Renyi converges to von Neumann entropy when $\alpha$ tends to $1$,
\begin{equation}
\lim_{\alpha\rightarrow 1}S_\alpha(\rho)=S(\rho).
\end{equation}
The application of Theorem \ref{thm:main} to the entanglement rate problem for Renyi entropy is discussed in the next section.

\section{Entropy rate for Renyi entropy}\label{sec:Renyi}

In this section we return to  a system $aABb$ discussed in Section \ref{sec:SIE} and measure the entanglement in it with the entanglement Renyi entropy by measuring it on one part of the system
$$ E(\rho)=S_\alpha(\rho_{aA}),$$
for $\alpha>0$. Since the state of the whole system plus ancillas stays pure during the Hamiltonian evolution, entanglement Renyi entropy can be measured either from Ailice's side or from  Bob's side: $S_\alpha(\rho_{aA})=S_\alpha(\rho_{Bb}).$

Suppose that $d_B\leq d_A$.
The entanglement rate is given by 
$$\Gamma_\alpha=\frac{\alpha}{1-\alpha}\frac{1}{\Tr(\rho_{aA}^\alpha)}\Tr (H[\rho_{aAB}, \rho_{aA}^{\alpha-1}]).$$
Denoting $X=\rho_{aAB}/d_B^2$ and $Y=\rho_{aA}\otimes I_B/d_B$ the entanglement rate can be written as
$$\Gamma_\alpha=\frac{\alpha}{1-\alpha}d_B \frac{1}{\Tr(Y^\alpha)}\Tr (H[X, Y^{\alpha-1}]),$$
for $\Tr(X)=p=\frac{1}{d_B^2}$, $\Tr(Y)=1$, $0\leq X\leq Y\leq I$, and $-I\leq H\leq I$.

We are going to consider the case when $\alpha\neq 1$, since it has already been investigated previously, as discussed in Section \ref{sec:prelim}.
For $\alpha>1$, $\Tr(Y^\alpha)\leq \Tr(Y)=1$, therefore we cannot eliminate it from the inequality
\begin{equation}\label{eq:Renyi_ent-2}
\Gamma_\alpha\leq \frac{\alpha}{|1-\alpha|}d_B \frac{1}{\Tr(Y^\alpha)}\|[X, Y^{\alpha-1}]\|_1\ .
\end{equation}

Using the explicit form of the Theorem \ref{thm:main} for the power function \ref{power-thm1}-\ref{power-thm2} the entanglement rate is bounded above by
\begin{itemize}
\item $2<\alpha$:  for large enough $d_B$, see (\ref{eq:bound-p}), $ \Gamma_\alpha\leq 9 \frac{\alpha}{\alpha-1}\frac{1}{\Tr(\rho_{aA}^\alpha)}\left(d_B^{\alpha-1}-\frac{1}{d_B^{\alpha-1}}\right)$,
\item $1<\alpha<2$: for large enough $d_B$, see (\ref{eq:bound-p}), $  \Gamma_\alpha\leq 9 \frac{\alpha}{\alpha-1}\frac{1}{\Tr(\rho_{aA}^\alpha)}(d_B^{\alpha+1}-d_B^{\alpha-1})(1-(1-\frac{1}{d_B^2})^{\alpha-1})$.
\end{itemize}

Note that these bounds are decreasing with dimension $d_B$, meaning that the rate of change of the entanglement in the system will slow down with the increase of the dimension. In other words, for sufficiently large dimension $d_B$ the entanglement will not change much in time if measured by Renyi entanglement entropy with $\alpha> 1$.

\section{Entropy rate for Tsallis entropy}\label{sec:Tsallis}

In this section we have a system $aABb$ discussed in Section \ref{sec:SIE} and measure the entanglement in it with the entanglement Tsallis entropy (\ref{eq:Tsallis}) by measuring it on one part of the system
$$E(\rho)=S_q (\rho_{aA}),$$
for $q>1$. Since the state of the whole system plus ancillas stays pure during the Hamiltonian evolution, entanglement Tsallis entropy can be measured either from Ailice's side or from  Bob's side: $S_q(\rho_{aA})=S_q(\rho_{Bb}).$

Suppose that $d_B\leq d_A$.
The entanglement rate is given by 
$$\Gamma_q=\frac{q}{1-q}\Tr (H[\rho_{aAB}, \rho_{aA}^{q-1}]).$$
Denoting $X=\rho_{aAB}/d_B^2$ and $Y=\rho_{aA}\otimes I_B/d_B$ the entanglement rate can be written as
$$\Gamma_q=\frac{q}{1-q}d_B^{q+1}\,\Tr (H[X, Y^{q-1}]),$$
for $\Tr(X)=p=\frac{1}{d_B^2}$, $\Tr(Y)=1$, $0\leq X\leq Y\leq I$, and $-I\leq H\leq I$. 

Using the explicit form of the Theorem \ref{thm:main} for the power function \ref{power-thm1}-\ref{power-thm2} the entanglement rate is bounded above by
\begin{itemize}
\item $2\leq q$:  for large enough $d_B$, see (\ref{eq:bound-p}), $\Gamma_q\leq 9 \frac{q}{q-1} \left(d_B^{q-1}-\frac{1}{d_B^{q-1}}\right)$,
\item $1<q<2$: for large enough $d_B$, see (\ref{eq:bound-p}),   $\Gamma_q\leq 9 \frac{q}{q-1}d_B^{q+1}\left(1-\frac{1}{d_B^2}\right)\left(1-(1-\frac{1}{d_B^2})^{q-1}\right)$.
\end{itemize}

Note that for $q>1$ these bounds are increasing with dimension $d_B$, which means that the rate of change of entanglement will grow in the system when measured by Tsallis-$q$ entanglement entropy. 

\section{Proof of Theorem \ref{thm:main}}\label{sec:Proof}
We will prove the following upper bound
$$\|[X, f(Y)])\|_1\leq 9\, p (f(1)-f(p)).$$
The second part in (\ref{eq:thm-main}) is derived from this one by substituting $X$ with $Y-X$.

\begin{proof}
The essence of the proof follows \cite{AMV13}. For completeness we provide the proof with all details. Since the trace-norm can be written as the maximum
$$\ \|[X, f(Y)])\|_1=\max_{0\leq H\leq I}2\left|\Tr (H[X, f(Y)])\right|$$
we will focus our attention on the expression
$$\Lambda(H)=2\left|\Tr (H[X, f(Y)])\right| $$
for $0\leq H\leq I$ and conditions on $X$ and $Y$ stated in the Theorem \ref{thm:main}.

Consider the eigendecomposition of the operator $Y$ with ordered eigenvalues
$$ Y=\sum_j y_j \ket{\phi_j}\bra{\phi_j}, \qquad \text{with } 0<y_N\leq \dots \leq y_1\leq 1.$$
Denote the corresponding matrix elements of operators $X$ and $H$ as follows
$$X=\sum_{i,j} x_{i,j} \ket{\phi_i}\bra{\phi_j}, \qquad H=\sum_{i,j} h_{i,j} \ket{\phi_i}\bra{\phi_j} .$$
Since $f$ is monotonically increasing, reorder and separate eigenvalues of $Y$ in the following way:\begin{align}
&f(1)>f(y_{j_1}) \geq f(p), \ \text{ for } j_{i_1}\in I_0=\{1,\dots, n_1\}\\
&f(p)>f(y_{j_{2}})\geq 2f(p)-f(1), \ \text{ for } j_{i_2}\in I_1=\{n_1+1,\dots, n_2\}\\
&\dots\\
&(k-1)f(p)-(k-2)f(1)>f(y_{i_k})>{f(0)\geq} kf(p)-(k-1)f(1), \ \text{ for } j_{i_k}\in I_{k-1}=\{n_k+1,\dots, n_{k+1}=N\}.
\end{align}
Then we have
\begin{align}
\Lambda(H)&=2\left|\Tr (H[X, f(Y)])\right|\\
&=2\left|\sum_{i<j}(f(y_i)-f(y_j)) (x_{ij}h_{ji}-x_{ji}h_{ij} )\right|.
\end{align}
The sum $\Lambda(H)=\sum_{i<j}$ can only be split into the following tree terms: $$\sum_{i<j}=\sum_{m=0}^{k-2}\sum_{\substack{i<j \\ i,j\in I_m\cup I_{m+1}}}+\sum_{m=0}^{k-3}\sum_{i\in I_m}\sum_{l\geq m+2}\sum_{j\in I_l}-\sum_{m=0}^{k-1}\sum_{\substack{i<j\\ i,j\in I_m}}.$$
The first sum runs over indices in two adjoint intervals.  The second sum runs over indices that are separated by at least one interval. The third sum runs over indices in one interval.
Denoting
\begin{align}
\Lambda_1&=2\left| \sum_m\sum_{\substack{i<j \\ i,j\in I_m\cup I_{m+1}}}(f(y_i)-f(y_j)) (x_{ij}h_{ji}-x_{ji}h_{ij} )\right|\\
\Lambda_2&=2\left|\sum_m\sum_{i\in I_m}\sum_{l\geq m+2}\sum_{j\in I_l}(f(y_i)-f(y_j)) (x_{ij}h_{ji}-x_{ji}h_{ij} )\right|\\
\Lambda_3&=2 \left| \sum_{m=1}^{k-1}\sum_{\substack{i<j\\ i,j\in I_m}}(f(y_i)-f(y_j)) (x_{ij}h_{ji}-x_{ji}h_{ij} )\right|.
\end{align}
we get 
\begin{align}
\Lambda(H)&\leq \Lambda_1+\Lambda_2+\Lambda_3.\label{eq:L}
\end{align}

Consider the sum appearing in $\Lambda_1$. For a fixed $m$ restrict all operators $X, Y$ and $H$ on a space generated by eigenvalues $y_j$ with $j\in I_m\cup I_{m+1}$ and denote them with a tilde above. The indices $i$ and $j$ are run over two adjoint intervals, therefore $f(y_j)-f(y_i)\leq 2 (f(1)-f(p)).$ Then 
\begin{align}
2\left|\Tr (\tilde{H}[\tilde{X}, f(\tilde{Y})])\right|&\leq\|[\tilde{X}, f(\tilde{Y})]\|_1\\
&=\|[\tilde{X}, f(\tilde{Y})-f(\tilde{y}_{min})]\|_1\\
&\leq \|\tilde{X}\|_1 \|f(\tilde{Y})-f(\tilde{y}_{min}) \|\\
&= \Tr \tilde{X} \ \|f(\tilde{Y})-f(\tilde{y}_{min}) \|\\
&= \Tr \tilde{X} \ \|f(\tilde{y}_{max})-f(\tilde{y}_{min}) \|\\
&= (p_m+p_{m+1}) \ \|f(\tilde{y}_{max})-f(\tilde{y}_{min}) \|\\
&\leq (p_m+p_{m+1}) 2(f(1)-f(p)),
\end{align}
where $\Tr({X}|_{I_m})=p_m$, i.e. this is a trace of the operator $X$ restricted to the space generated by eigenvalues $y_j$ such that $j\in I_m$. Note that $\sum_{m=0}^{k-1}p_m=p=\Tr(X).$
Therefore the whole sum $\sum_m\sum_{i,j\in I_m\cup I_{m+1}}$ is bounded above by 
\begin{align}\label{eq:L1}
\Lambda_1&\leq  4p(f(1)-f(p)).
\end{align}

Similarly for the third sum that runs over one interval we have, $f(y_j)-f(y_i)\leq  (f(1)-f(p)),$ and
\begin{align}
\Lambda_3&\leq \Tr(X|_{I_m}) \ \|f(\tilde{y}_{max})-f(\tilde{y}_{min}) \|\nonumber\\
&= p_m \ \|f(\tilde{y}_{max})-f(\tilde{y}_{min}) \|\nonumber\\
&\leq  p(f(1)-f(p)).\label{eq:L3}
\end{align}

In the sum appearing in $\Lambda_2$ the indices are separated by at least one interval, therefore
$$f(y_i)-f(y_j)>f(1)-f(p).$$
For such points from Assumption \ref{Assum:func} we have that 
$$ y_i^{-1/2}y_j^{1/2}(f(y_i)-f(y_j)) \leq p^{1/2}(f(1)-f(p)).$$
From trivial manipulations we have 
\begin{align}
\frac{1}{2}\Lambda_2\leq&\left| \sum_{i<j} \left\{ (f(y_i)-f(y_j))^{1/2}{y_i}^{-1/4}{y_j}^{-1/4} x_{ij} \right\}\right|\left|\left\{ (f(y_i)-f(y_j))^{1/2}{y_i}^{1/4} {y_j}^{-1/4} h_{ji} \right\}\right|\\
&+\left|\sum_{i<j} \left\{ (f(y_i)-f(y_j))^{1/2}{y_i}^{-1/4}{y_j}^{-1/4} x_{ji} \right\}\right|\left|\left\{ (f(y_i)-f(y_j))^{1/2}{y_i}^{1/4}{y_j}^{1/4} h_{ij} \right\}\right|.
 \end{align}
Using Cauchy-Schwartz inequality we obtain
 \begin{align}
\frac{1}{2}\Lambda_2 \leq & 2 \left\{ \sum_{i<j} (f(y_i)-f(y_j)){y_i}^{-1/2}{y_j}^{-1/2} x_{ij}x_{ji} \right\}^{1/2} \left\{ \sum_{i<j}(f(y_i)-f(y_j)){y_i}^{1/2}{y_j}^{1/2} h_{ji}h_{ij} \right\}^{1/2}\\
\leq &2 p^{1/2}(f(1)-f(p))  \left\{ \sum_{i,j} {y_j}^{-1} x_{ij}x_{ji} \right\}^{1/2} \left\{ \sum_{i,j} y_i h_{ji}h_{ij} \right\}^{1/2}\\
\leq &2 p^{1/2}(f(1)-f(p))  \left\{ \sum_{j} {y_j}^{-1} (X^2)_{jj} \right\}^{1/2}\\
 \leq &2 p(f(1)-f(p)).
 \end{align}
Here for the second inequality we used that $f$ satisfies Assumption \ref{Assum:func}. The third inequality follows from the fact that $0<H<I$, and therefore $$ \sum_{i,j} y_i h_{ji}h_{ij}=\sum_j y_i(H^2)_{ii}=\Tr (YH^2)\leq \Tr (Y)=1.$$ The last inequality follows from the fact that $0\leq X\leq Y\leq 1$, therefore $$ \sum_{j} y_j^{-1} (X^2)_{jj}=\Tr (XY^{-1}X)\leq \Tr (X)=p.$$
Therefore, $$\Lambda _2\leq 4 p(f(1)-f(p)).$$ 

Summing this bound with (\ref{eq:L1}) and (\ref{eq:L3}), from (\ref{eq:L}) we obtain
$$\Lambda(H) \leq  9 p(f(1)-f(p)).$$
Since the right-hand side does not depend on $H$, the statement in Theorem \ref{thm:main} is derived.
\end{proof}

\section{Power function}\label{sec:Func}

Consider $f(t)=t^{\beta}$ for $\beta>0$.  The function $g$ defined in (\ref{eq:monotone-function}) takes the following form
$$g(y)= (y^\beta+1-p^\beta)^{-1/2\beta} y^{1/2} (1-y^\beta).$$
Recall that from (\ref{eq:ass1}), the monotonicity of $f$ and the condition $0<y<x\leq 1$, we have $y<p$. Proving that for small $p$ this function is monotonically increasing will prove the inequality (\ref{eq:ass2}). In other words, for $y<p$ we need to prove that
$$g(y)\leq g(p).$$
Taking this inequality to the power $2\beta>0$ it is equivalent to
$$ (y^\beta+1-p^\beta)^{-1} y^{\beta} (1-y^\beta)^{2\beta}< p^{\beta}(1-p^\beta)^{2\beta}.$$
Rename $t:=y^\beta$ and $q:=p^\beta$, then since $y<p$, we have $t<q$, and the above inequality is equivalent to the monotonicity statement of the following function
$$ h_\beta(t):=t(t+1-q)^{-1} (1-t)^{2\beta}< q(1-q)^{2\beta}=h_\beta(q).$$
The function $h_\beta$ is non-negative on the interval $[0,1]$ and has only zeros at $0$ and $1$. 
Consider its derivative
\begin{align}
h_\beta'(t)=-(t+1-q)^{-2}(1-t)^{2\beta-1}\Bigg\{2\beta t^2 +(2\beta +1)(1-q)t-(1-q) \Bigg\}.
\end{align}
First two terms are non-negative for all $0<t<1$, $0<q<1$. The second term has two zeros, one at some $t<0$ and another one at $t_0$ lies on the interval $0<t<1$, where
$$t_0=\frac{1}{4\beta}\left \{\sqrt{(2\beta+1)^2(1-q)^2+8\beta(1-q)}-(2\beta+1)(1-q) \right\}.$$
Therefore, the function $h_\beta$ is monotonically increasing on $[0, t_0]$ and monotonically decreasing on $[t_0, 1]$. 

Since $t<q$, we aim to show that there are small enough $q$ such that $q<t_0$, which proves that function $h_\beta(t)$ increasing monotonically for such small $q$ and $t<q$.  The following inequalities are equivalent
\begin{align}
q&<t_0\\
(2\beta+1)+(2\beta-1)q&<\sqrt{(2\beta+1)^2(1-q)^2+8\beta(1-q)}\\
0&<8\beta q^2-16\beta(\beta+1)q+8\beta\\
0&<q^2-2(\beta+1)q+1.
\end{align}
Note that on the second line the left-hand side is positive for all $\beta>0$ and $q\leq 1$.

For small enough $q$, i.e.
$$q<\beta+1-\sqrt{(\beta+1)^2-1}=\gamma(\beta), $$
 we have $q<t_0$.  (Note that $0<\gamma(\beta)<1$ for all $\beta$.)
 
 Therefore since the function $h_\beta$ is monotonically increasing on $[0, t_0]$ 
$$\text{for }t< q\leq t_0,\text{ we have }h_\beta(t)< h_\beta(q).$$
The Assumption \ref{Assum:func} is proved for the function $f(t)=t^\beta$, $\beta>0$ and 
\begin{equation}\label{eq:bound-p}
p<\left(\beta+1-\sqrt{(\beta+1)^2-1}=\gamma(\beta) \right)^{1/\beta}.
\end{equation}

\section{Conclusion}

The main result of this paper, Theorem \ref{thm:main}, provides an upper bound on the trace-norm of the commutator between two operators, which can be seen as an entropy rate for a large class of entropy functions, e.g. (\ref{eq:ent-measure}) and (\ref{eq:ent-measure2}). This class of entropy measures include two generalizations of von Neumann entropy: Renyi and Tsallis entropies. Theorem \ref{thm:main} is a true generalization of the result for the von Neumann entropy (\ref{eq:SIM2}), first derived in \cite{AMV13}, as it precisely recovers this bound for the logarithmic function.

The trace-norm of the commutator considered in the main Theorem \ref{thm:main} was first explored in \cite{AK12} by Audenaert and Kittaneh, where they conjectured another upper bound on it (\ref{eq:Aud-conj}). While it is impossible to compare these bounds for a general function, the upper bound presented in this paper provides a sharper scaling comparing to the conjectured one for a logarithmic function and for power function with the power no greater than 1, for a power greater than 1 the conjectured bound would give a better scaling, if proved. Theorem \ref{thm:main} also enlarges the class of function for which the upper bound is found, e.g. it includes power function with a power less than -1.

The constant $9$, which appears in the upper bound in Theorem \ref{thm:main}, is not optimal. For von Neumann entropy Audenaert \cite{A13} provided a completely different proof of SIM with constant $2$. Numerical evidence \cite{B07} suggests that the optimal constant in SIM for von Neumann entropy is $1$, although no proof is known for it at the time of writing this paper. Unfortunately, Audenaert's proof can not be easily generalized to functions other than logarithmic, so in this paper we have sacrificed the constant but generalized the problem to a large class of functions.

Taking a power function in Theorem \ref{thm:main} we derive an upper bond on the entanglement rate with entanglement Renyi entropy taken as a measure of entanglement in the system. Similarly, directly applying Theorem \ref{thm:main} one obtains the upper bound on the entanglement rate for Tsallis entropy.

Our result holds for closed systems evolving under unitary Hamiltonian evolution. The entanglement rate problem in open systems was first considered in \cite{V15}, with the relative entropy of entanglement and a quantum mutual information  taken as entanglement measures. The derivation of the upper bound on the entanglement rate relied on the following upper bound similar to SIM
$$\left|\Tr(L^\dagger[LX, \ln(Y)]) \right| \leq c\|L\|^2 p\ln(1/p),$$
with the same conditions of $X,Y$, and $L$ being a bounded linear operator.
We expect a similar generalization to hold for this expression, leading to
\begin{equation}\label{eq:general_SIM_L}
\left|\Tr(L^\dagger[LX, f(Y)]) \right| \leq c\|L\|^2 p (f(1)-f(p)).
\end{equation}
We do not provide a complete proof of this statement here, because it is {not clear for which reasonable entanglement measures} it will be useful, provided this statement is valid. But we expect the proof of (\ref{eq:general_SIM_L}) to follow the same lines as the proof of Theorem \ref{thm:main} and the modifications presented in \cite{V15}.

Problems that are still open in the spirit of this generalization include increasing the class of function for which generalized SIM and SIE hold, proving SIE in open systems for a general entanglement measure, proving or disproving conjecture (\ref{eq:Aud-conj}) posed by Audenaert and Kittaneh.

\vspace{0.3in}
\textbf{Acknowledgments.} A.V. is grateful to Elliott H. Lieb for hosting her visit to Princeton University and engaging in productive discussions. A.V. thankful to Zhengyan Shi for careful reading of the paper and making valuable corrections.

\end{document}